\documentstyle[preprint,aps,version2]{revtex}
\newcommand{\dfr}[2]{ \displaystyle\frac{#1}{#2} }
\newcommand{\ubu}[1]{\:\bar{\cal U}_{v,s_1'}(k_1')\:{#1}\:{\cal
U}_{v,s_1}(k_1)\;}
\newcommand{\vbv}[1]{\:\bar{\cal V}_{v,s_2}(k_2)\:{#1}\:{\cal
V}_{v,s_2'}(k_2')}
\newcommand{\rdv}{\;\rlap/{\!\!{\Delta}\! v}}
\begin{document}
\vspace{-20ex}
\begin{flushright}
\vspace{-2.0mm}
       \it{TUIMP-TH-93/54} \\
       \it{CCAST-93-37} \\
\vspace{-2.0mm}
       \it{Nov., 1993}\\
\vspace{5.0ex}
\end{flushright}

\draft
\begin{title} {\bf
General relations of heavy quark-antiquark potentials \\
induced by reparameterization invariance\footnote{Work
supported by the National Natural Science Foundation of China. } }
\end{title}
\author{ Yu-Qi Chen$^a$ ~~~ and ~~~ Yu-Ping Kuang$^{a,b,c}$ }
\begin{instit}
$^a$China Center of Advanced Science and Technology (World Laboratory),\\
P. O. Box 8730, Beijing 100080, China\\
$^{b}$Institute of Modern Physics, Tsinghua University, Beijing 100084,
         China\footnote{Mailing address.}\\
$^{c}$Institute of Theoretical Physics, Academia Sinica, Beijing 100080, China
\end{instit}

\vspace{1.5cm}
\centerline{ Abstract}
\vspace{-1.cm}

\begin{abstract}
A set of general relations between the spin-independent and spin-dependent
potentials of heavy quark and anti-quark interactions are derived from
reparameterization invariance in the Heavy Quark Effective Theory.
It covers the Gromes relation and includes some new interesting relations
which are useful in understanding the spin-independent and spin-dependent
relativistic corrections to the leading order nonrelativistic potential.
\end{abstract}

\narrowtext
\newpage

spin-independent (SI) and spin-dependent (SD) parts, is of special interest
in heavy quark physics because it provides useful knowledge bridging the
first principles of QCD and the experimental data in the $c\bar{c}$ and
$b\bar{b}$ systems in which nonperturbative QCD effects are significant.
There have been two kinds of approaches to the SD potentials in the
literatures. The first kind of  approach is in the framework of $1/m$
expansion, where $m$ stands for the heavy-quark mass. The formulae for the
SD potentials in this approach were first given by Eichten and
Feinberg\cite{1}, in which the potentials were  expressed in terms of
certain correlation functions of color-electric and color-magnetic fields
weighted by a Wilson loop factor. Assuming color-electric confinement, they
evaluated the correlation functions containing color-magnetic fields to
leading order in QCD perturbation.  Based on an intuitive color-electric flux
tube picture of color-electric confinement, Buchm\"{u}ller\cite{20} pointed
out that the long-range interaction in the spin-orbit coupling potential
should be of opposite sign compared with the above evaluation.  Later
Gromes\cite{2} derived an important relation between the SI and SD potentials
from the Lorentz invariance of the total potential and the correlation function
expressions given in Ref.\cite{1}. This relation is more fundamental, and it
implies that the simple short-range assumption of the correlation functions
containing color-magnetic fields is not always adequate. It is shown that
Buchm\"{u}er's result is consistent with this relation. The other kind of
approach is to calculate loop contributions to the SD potentials in  the
full perturbative QCD theory without making $1/m$ expansion\cite{3,4}.
In this kind of approach, logarithmic $m$-dependence of the potentials
emerges from the loop diagrams. Pantaleone, Tye, and Ng\cite{4} pointed
out that Gromes relation is still satisfied by the corresponding potentials
in this approach to 1-loop level, however, there are extra $\ln m$-dependent
spin-orbit coupling terms which are not included in the first kind of approach.
It will be interesting if one can understand the potentials from a
more general stand point which can cover both of these two approaches.

In this paper, we shall show  that the reparameterization
invariance\cite{14,15,16,17} in the Heavy Quark Effective Theory\cite{6}
(HQET) leads to a set of general relations between the SI and SD potentials
which do cover the two approaches. We shall see that the Gromes relation
is included in these relations, and there is a new relation  between the
SI and SD potentials which covers the property of the extra spin-orbit
coupling potentials in the second approach. Furthermore, there are two
more new relations between the SI potentials of different orders in the
$1/m$ expansion, which are useful for understanding the spin-independent
relativistic corrections to the leading order static potential.

The  conventional way of studying the interaction potential between heavy
quark and antiquark is to extract it from the scattering amplitude.
Consider a heavy quark $Q_1$ with mass $m_1$ and a heavy antiquark
$\bar{Q}_2$ with mass $m_2$. Let  $p_1$ ($p_2$) and $p_1'$ ($p_2'$) be the
initial- and final-state momenta of $Q_1$ and $\bar{Q}_2$, respectively.
The on-shell conditions are
\begin{equation}
p_1^2=p_1^{'2}=m_1^2,  ~~~~~~~ p_2^2=p_2^{'2}=m_2^2.
\end{equation}

Let $v$ be a four-velocity satisfying $v^2=1$, which can be $v=(1,0,0,0)$
but not necessarily. As what is usually done in the HQET, we parameterize
the momenta $p_1$, $p_2$, $p_1'$, $p_2'$ by
\begin{equation}
\begin{array}{l}
p_{1}=m_1v+k_{1}, ~~~~~~~ p_{2}=m_2v+k_{2},\\
p_{1}'=m_1v+k_{1}', ~~~~~~~ p_{2}'=m_2v+k_{2}',
\label{pp}
\end{array}
\end{equation}
where $k_1$, $k_1'$, $k_2$, and $k_2'$ are residual momenta. We consider
here a nonrelativistic scattering process in which $k_i,k_i' \ll m_iv$
($i=1,2$), so that $1/m$ expansion makes sense. Note that the momentum
transfer
\begin{equation}
q\equiv p_1'-p_1=p_2-p_2'
\label{3}
\end{equation}
is independent of the parameter $v$. In the HQET, the heavy quark field
$h_{v+}(x)$ and the heavy antiquark field $h'_{v-}(x)$ are related to the
heavy quark field $\psi(x)$ in the full theory by\cite{6}
\begin{equation}
\begin{array}{lcl}
 h_{v+}(x) &\equiv& P_{+}h_{v}(x) = P_+ e^{imv\cdot x}\psi (x),\\
 h'_{v-}(x) &\equiv& P_{-}h'_{v}(x) =P_-e^{-imv\cdot x}\psi (x),
\end{array}
\end{equation}
where $P_{\pm}\equiv \dfr{1\pm \rlap/{v}}{2}$. Let $| v,+,k,s\rangle$
and $| v,-,k,s\rangle$ be, respectively, the state-vectors for a heavy
quark and an antiquark in the Hilbert space.
 The wave functions ${\cal U}_{v,s}(k)$ and ${\cal V}_{v,s}(k)$
for the heavy quark and antiquark are defined by\cite{6}
\begin{equation}
\begin{array}{lcl}
\langle 0|h_{v+}(x)|v,+,k,s\rangle  &=&\sqrt{\dfr{m}{E}}{\cal U}_{v,s}(k)
e^{-ik\cdot x},\\
\langle 0|h'_{v-}(x)|v,-,k,s\rangle  &=&\sqrt{\dfr{m}{E}}{\cal V}_{v,s}(k)
e^{ik\cdot x},
\end{array}
\end{equation}
and the heavy quark polarization vector $s_v$ in the $v$ parameterized
effective
theory is related to the spin vector $s$ by\cite{15,17}
\begin{equation}
s_v^\mu=s^\mu-\dfr{p^\mu+mv^\mu} {m+p\cdot v}\;s\cdot v.
\label{pol}
\end{equation}

There are three independent momenta among $p_1$, $p_1'$, $p_2$, $p_2'$
due to momentum conservation. Introduce
\begin{equation}
\kappa_1\equiv \dfr{k_1+k_1'}{2}, ~~~~~~~~~
\kappa_2\equiv \dfr{k_2+k_2'}{2}.
\end{equation}
It is convenient to use $\kappa_1$, $\kappa_2$, and $q$ (cf.(\ref{3}))
as three independent momentum parameters. To order-$1/m^2$, the general
form of the Lorentz- and C-, P-, T- invariant Bethe-Salpeter irreducible
scattering amplitude\cite{3}  can be written (in the Pauli form) as
\begin{equation}
\begin{array}{lcl}
&& A_{s'_1s_1s'_2s_2}(v;k_1',k_1,k'_2,k_2) \\
  &=& \left[\;U_0(q)
+\dfr{1}{m_1}\, U_1(q)v\cdot \kappa_1
+\dfr{1}{m_2}\, U_2(q)v\cdot \kappa_2
+\dfr{1}{m_1^2} \,U_3(q)\, \kappa_1^2 \right.\\[2mm]
&+& \left. \dfr{1}{m_1m_2} \,U_4(q)\, \kappa_1\cdot \kappa_2
+\dfr{1}{m_2^2} \,U_5(q)\, \kappa_2^2 \;\right]
\, \ubu{}\vbv{}\\[2mm]
&-&\dfr{i}{ m_1^2} U_6(q)\;
 \ubu{\sigma_{\mu\nu} q^\mu \kappa_1^\nu }\vbv{}\\[2mm]
&-&\dfr{i}{ m_2^2} U_7(q)\;
 \ubu{} \vbv{\sigma_{\mu\nu}q^\mu \kappa_2^\nu} \\[2mm]
&+&\dfr{i}{ m_1m_2} U_8(q)\;
 \ubu{\sigma_{\mu\nu} q^\mu \kappa_2^\nu } \vbv{}\\[2mm]
&+&\dfr{i}{ m_1m_2} U_9(q)\;
 \ubu{} \vbv{\sigma_{\mu\nu}q^\mu \kappa_1^\nu } \\[2mm]
&+&\dfr{1}{2m_1m_2} [\; U_{10}(q)\,q^2+U_{11}(q)\,\kappa_1\cdot\kappa_2\;]\;
 \ubu{\sigma^{\mu\nu} } \vbv{\sigma_{\mu\nu} }\\[2mm]
&-&\dfr{1}{ m_1m_2} U_{12}(q)\;
 \ubu{ \sigma^{\mu\nu}q_\mu } \vbv{ \sigma_{\lambda\nu}q^\lambda }\\[2mm]
&+&   \dfr{1}{ m_1m_2} U_{13}(q) \ubu{\sigma_{\mu\nu} \kappa_1^\mu}
\vbv{\sigma^{\lambda\nu}\kappa_{2\lambda}}\\[2mm]
&+&   \dfr{1}{ m_1m_2} U_{14}(q)
\ubu{\sigma_{\mu\nu}\kappa_2^\mu}\vbv{\sigma^{\lambda\nu}\kappa_{1\lambda}},\\[2mm]
\end{array}
\label{mpot}
\end{equation}
where $U_0(q)-U_5(q)$ are SI potentials, and $U_6(q)-U_{14}(q)$ are SD
potentials. We have explicitly extracted the heavy quark mass dependence
in each term according to the following rules: (a) for the SI terms, every
small momentum $\kappa_i$ is associated with a factor $1/m_i$; (b) for the
SD terms, every $\kappa_i$ is associated with a factor $1/m_i$, while every
$\sigma_{\mu\nu}q^\mu$ is associated with a factor $1/m_1$, or $1/m_2$
depending on whether $\sigma_{\mu\nu}q^\mu$ is sandwiched in the quark wave
functions or the antiquark wave functions. These rules are consistent with
the spirit of $1/m$ expansion in the HQET. In general, the above rules may
be invalid in the following cases:

i) the fundermental or the effective theory contains an unusual large tensor
interaction which is not suppressed by a factor of $1/m$\cite{11};

ii) the fundermental theory contains an axial-vector couplings which is
not suppressed by a factor of $1/m$\cite{11}. (in this case,
$P_\pm\gamma_5\gamma_\mu P_\pm= \pm P_{\pm}i\gamma_5\sigma_{\mu\nu}v^\nu P_\pm
=\pm\dfr{i}{2} P_\pm \epsilon_{\mu\nu\rho\sigma}\sigma^{\rho\sigma}v^\nu P_\pm
\equiv \pm P_\pm \sigma^v_{\mu}P_\pm$, where $\sigma^v_{\mu}$ is twice of
the spin operator in the HQET so that there is a leading order spin-spin
interaction which is not $1/m$-suppressed);

iii)  the coupling is proportional to the heavy quark mass like the Yukawa
coupling in the electroweak theory;

iv) theories with pseudoscalar particle exchange, which  does not have leading
order static potentials\cite{11}.

In QCD none of the above cases happens, therefore these rules do apply
to (\ref{mpot}). Note that there is no order-$1/m$ SD terms in (\ref{mpot}).
This is due to the identities
\begin{equation}
P_\pm\sigma_{\mu\nu}v^\mu P_\pm=0,~~~~~~~~~P_\pm\gamma_\mu P_\pm =\pm v_\mu
\end{equation}
in the HQET.

As we have mentioned, the form (\ref{pp}) is only a kind of parameterization
of the momenta $p_1$, $p_2$, $p_1'$, $p_2'$.  For given $p_1$, $p_2$,
$p_1'$, and $p_2'$, different sets of $v$ and $k_i$, $k'_i$ correspond to
different parameterizations which should give the same physical predictions.
In other words, (\ref{mpot}) should be invariant against the change of the
parameterization. This is the so-called {\it reparameterization invariance}
in the HQET, and it has proved to be very powerful in the study of heavy
flavor physics\cite{14,15,16,17}. Let us consider an infinitesimal  change
of $v$ in (\ref{pp}),
\begin{equation}
v\to v'=v+\Delta v.
\label{dv}
\end{equation}
The constraint $v^2=1$  implies that
\begin{equation}
\Delta v\cdot v=0.
\end{equation}
To keep the physical momenta $p_1$, $p_2$, $p_1'$, and  $p_2'$ unchanged, the
corresponding changes of $k_i$ and $k_i'$ should be
\begin{equation}
\Delta k_{1}= \Delta k_{1}'= - {m_1}\, {\Delta v}, ~~~~~~
\Delta k_{2}= \Delta k_{2}'= - {m_2}\, {\Delta v}.
\label{dk}
\end{equation}

The general form of the infinitesimal change of the heavy quark
wave functions corresponding to (\ref{dv})-(\ref{dk})
has been given in Ref.\cite{17}.  To the order-$1/m$, it is\cite{15,17}
\begin{equation}
\begin{array}{lcl}
\Delta {\cal U}_{v,s}(k) &=&
\dfr{\rdv }{2}\left( 1+ \dfr{ \rlap/ k }{2m} \right) {\cal U}_{v,s}(k) \\[2mm]
\Delta {\cal V}_{v,s}(k) &=&
-\dfr{\rdv }{2}\left( 1- \dfr{ \rlap/ k }{2m} \right) {\cal V}_{v,s}(k)
\label{dw}
\end{array}\end{equation}

The potentials $U_0(q),\cdots,U_{14}(q)$ in (8) are physical quantities which
should be invariant under the reparameterization transformation
(\ref{dv})-(\ref{dk}). Thus, the infinitesimal of the scattering amplitude
$A_{s'_1s_1s_2's_2}(v;k_1',k_1,k_2',k_2)$ can be easily worked out from
(\ref{dv})-(\ref{dw}). To order-$1/m$, it reads
\begin{equation}\begin{array}{lcl}
&& \Delta A_{s_1's_1s_2's_2 }(v;k_1',k_1,k_2',k_2) \\[2mm]
&=&  \left\{
\left[ U_0(q)+2U_1(q)-4U_3(q)-2U_4(q)\right]\dfr{\kappa_1\cdot\Delta{v}}{2m_1}
 + \left[ U_0(q)+2U_2(q)- 4U_5(q)
 \right.\right. \\[2mm]
&-& \left.\left. 2U_4(q)\right] \dfr{\kappa_2\cdot\Delta{v}}{2m_2}
\right\}  \ubu{} \vbv{} \\[2mm]
&+& \dfr{i}{4m_1} \left[ U_0(q) + 4U_6(q)- 4 U_8(q) \right]
 \ubu{\sigma_{\mu\nu} q^\mu \Delta\!v^\nu } \vbv{} \\[2mm]
&+& \dfr{i}{4m_2} \left[ U_0(q) + 4U_7(q)- 4 U_9(q) \right]
 \ubu{} \vbv{\sigma_{\mu\nu} q^\mu \Delta\!v^\nu } \\[2mm]
&-& \left(\dfr{\kappa_1\cdot\Delta{v}}{2m_1} +
          \dfr{\kappa_2\cdot\Delta{v}}{2m_2} \right) U_{11}(q)
 \ubu{\sigma_{\mu\nu} } \vbv{\sigma^{\mu \nu} }\\[2mm]
&-& \dfr{1}{m_2}\, U_{13}(q) \ubu{\sigma_{\mu\nu}\Delta\!v^\mu}
        \vbv{\sigma^{\lambda\nu}\kappa_{2\lambda}}\\[2mm]
&-&  \dfr{1}{ m_1} U_{13}(q)  \ubu{\sigma_{\mu\nu} \kappa_1^\mu}
        \vbv{\sigma^{\lambda\nu}\Delta\!v^\lambda}\\[2mm]
&-& \dfr{1}{m_1} U_{14}(q)  \ubu{\sigma_{\mu\nu}\Delta\!v^\mu}
\vbv{\sigma^{\lambda\nu}\kappa_{1\lambda}}\\[2mm]
&-&  \dfr{1}{ m_2} U_{14}(q)  \ubu{\sigma_{\mu\nu}\kappa_2^\mu }
\vbv{\sigma^{\lambda\nu}\Delta\!v^\lambda}.
\label{da}
\end{array}\end{equation}

For arbitrary $\kappa_1$ and  $\kappa_2$,
reparameterization invariance requires
\begin{eqnarray}
&U_0(q) +2 U_1(q) - 4 U_3(q) -2U_4(q) = 0,& \label{i1} \\
&U_0(q) +2 U_2(q) - 4 U_5(q) -2U_4(q) = 0,& \label{i2} \\
&U_0(q) +4 U_6(q) -4U_8(q) = 0,& \label{g1} \\
&U_0(q) +4 U_7(q) - 4 U_9(q)=0, &\label{g2} \\
&U_{11}(q)=0, ~~~~ U_{13}(q)=0,~~~~ U_{14}(q)=0. & \label{g4}
\end{eqnarray}
These are the general relations between the potentials in the momentum
representation that we obtain from the reparameterization invariance of
the scattering amplitude. We see that reparameterization invariance does
not give any constraints on $U_{10}(q)$ and $U_{12}(q)$.

 From the general symmetry argument, there can be terms containing
$U_{11}(q)$, $U_{13}(q)$, $U_{14}(q)$ in (\ref{mpot}). However, (\ref{g4})
shows that these terms are not consistent with reparameterization invariance,
and hence they should actually vanish. This explains why these terms never
appear in the results obtained fron specific dynamical
calculations\cite{1,20,2,3,4}.

Eqs.~(\ref{i1}) and (\ref{i2}) relate the leading order SI potential
$U_0(q)$ to its SI relativistic corrections $U_1(q)$, $U_2(q)$, $U_3(q)$,
$U_4(q)$, and  $U_5(q)$. These relations have never been shown in the
literatures. Although the two relations are still not enough to fix all
the five potentials $U_1(q)$-$U_5(q)$, they are at least helpful in
understanding the general properties of the SI relativistic corrections to
$U_0(q)$, which is one of the difficult problems in heavy quark physics.

Eqs.~(\ref{g1}) and (\ref{g2}) relate $U_0(q)$ to its SD relativistic
corrections. They are related to the Gromes relation. To see this precisely,
we make the Fourier transformation of Eq.~(\ref{da}) and derive the
relations between the potentials in the space-time representation corresponding
to (\ref{g1}) and  (\ref{g2}). Let $U_j(r)$ be the Fourier transform of
$U_j(q)$ with given $Q_1$-$\bar{Q}_2$ separation $r$. The relations
corresponding to (\ref{g1}) and (\ref{g2}) in terms of $U_j(r)$ are

\begin{eqnarray}
&\dfr{d}{dr}\,[\, U_0(r) +4 U_6(r) -4U_8(r) \,]\, =0, &\label{r1} \\
&\dfr{d}{dr}\,[\, U_0(r) +4 U_7(r) -4U_9(r) \,]\, =0. &\label{r2}
\end{eqnarray}

To compare this with the standard formulae given in Refs.\cite{1,20,2,3,4},
we also make the Fourier transformation of (\ref{mpot}). The relevant terms in
the space-time representation potential in (\ref{mpot}), when taking
$v=(1,0,0,0)$, are
\begin{equation}
\begin{array}{lcl}
V(r) &=& U_0(r)
+\left( \dfr{{\rm\bf{S}_1 }}{m_1^2}
+\dfr{{\rm\bf{S}_2 }}{m_2^2} \right) {\bf \cdot L}
\dfr{1}{r}\dfr{d}{dr}[\,U_6(r) +U_7(r) \,]\, \\[5mm]
&+&
\left( \dfr{{\rm\bf {S}_1 + {S}_2 }}{m_1m_2} \right)
{\bf \cdot L}
\dfr{1}{r}\dfr{d}{dr}[\,U_8(r) +U_9(r) \,] \\[5mm]
&+&
\dfr{4}{m_1m_2}\dfr{({\rm\bf {S}_1}\cdot {\rm\bf {r}})( {\rm\bf {S}_2}\cdot
{\rm\bf {r}})-\dfr{1}{3}{\rm\bf {S}_1} \cdot{\rm\bf {S}_2}\,r^2 }{r^2}
\bigtriangledown^2 U_{12}(r) \\[5mm]
&+&
\dfr{4}{m_1m_2}{\rm\bf{S}_1\cdot {S}_2}
\bigtriangledown^2\left[\,U_{10}(r)-\dfr{2}{3}U_{12}(r)\,\right] \\[5mm]
&+&
\left(\dfr{{\rm\bf{S}_1 }}{m_1^2}
-\dfr{{\rm\bf{S}_2 }}{m_2^2} \right) {\bf \cdot L}
\dfr{1}{r}\dfr{d}{dr}[\,U_6(r) -U_7(r) \,]\\[5mm]
&+&
 \left( \dfr{{\rm\bf {S}_1 - {S}_2 }}{m_1m_2} \right)
{\bf \cdot L}
\dfr{1}{r}\dfr{d}{dr}[\,U_8(r)- U_9(r) \,] +\cdots,
\end{array}
\label{cpot}
\end{equation}
where${\rm\bf{S}_1}$ and ${\rm\bf{S}_2}$ are, respectively, the spins
of $Q_1$ and $\bar{Q}_2$, and ${\rm\bf{L}}$ is the relative orbital
angular momentum. Let
\begin{equation}
\begin{array}{l}
V_0(r)\equiv U_0(r), ~~~~~~~~
V_1(r)\equiv U_6(r) + U_7(r) -\dfr{1}{2} U_0(r), \\
V_2(r)\equiv U_8(r) + U_9(r),  ~~~~~~~~
V_3(r)\equiv 4 U_{12}(r), \\
V_4(r)\equiv 12  \,\left[\,U_{10}(r)-\dfr{2}{3}U_{12}(r) \,\right],\\
V_5(r)\equiv U_6(r)- U_7(r), ~~~~~~~~
V_6(r)\equiv U_8(r) - U_9(r).
\end{array}
\end{equation}
In terms of $V_0(r),\cdots,V_6(r)$, (\ref{cpot}) can be written in the
standard form
\begin{equation}
\begin{array}{lcl}
V(r) &=& V_0(r)
+\left( \dfr{{\rm\bf{S}_1 }}{m_1^2}
+\dfr{{\rm\bf{S}_2 }}{m_2^2} \right) {\bf \cdot L}
\dfr{1}{r}\,\left[\,\dfr{1}{2}\dfr{dV_0(r)}{dr} + \dfr{dV_1(r)}{dr}\,\right]\;
+ \left( \dfr{{\rm\bf {S}_1 + {S}_2 }}{m_1m_2} \right)
{\bf \cdot L} \dfr{1}{r}\dfr{dV_2(r)}{dr}  \\[5mm]
&+&\dfr{1}{m_1m_2}\dfr{({\rm\bf {S}_1}\cdot {\rm\bf {r}})
( {\rm\bf {S}_2}\cdot
{\rm\bf {r}})-\dfr{1}{3}{\rm\bf {S}_1}
\cdot{\rm\bf {S}_2}\,r^2 }{r^2}\, \bigtriangledown^2V_{3}(r)
+\dfr{1}{3} \, \dfr{1}{m_1m_2} {\rm\bf{S}_1\cdot {S}_2}\bigtriangledown^2V_4(r)
\\[5mm] &+&
\left( \dfr{{\rm\bf{S}_1 }}{m_1^2}
-\dfr{{\rm\bf{S}_2 }}{m_2^2} \right) {\bf \cdot L}
\dfr{1}{r}\dfr{dV_5(r)}{dr}
+ \left( \dfr{{\rm\bf {S}_1 - {S}_2 }}{m_1m_2} \right)
{\bf \cdot L}
\dfr{1}{r} \dfr{dV_6(r)}{dr} +\cdots,
\end{array}
\label{vpot}
\end{equation}
which covers the forms given in Refs.\cite{1,20,2,3,4}. In terms of
$V_0(r),\cdots,V_6(r)$, the relations (\ref{r1}) (\ref{r2}) read
\begin{eqnarray}
&\dfr{d}{dr}\,[\, V_0(r)+ V_1(r)-V_2(r) +V_5(r)-V_6(r) \,]\, =0, &\\
&\dfr{d}{dr}\,[\, V_0(r)+ V_1(r)-V_2(r) -V_5(r)+V_6(r) \,]\, =0, &
\end{eqnarray}
and from which we obtain
\begin{eqnarray}
&\dfr{d}{dr}\,[\, V_0(r)+ V_1(r)-V_2(r) \,]\, =0, &\label{v1}\\
&\dfr{d V_5(r)}{dr}  = \dfr{d V_6(r)}{dr}.  &\label{v2}
\end{eqnarray}
Eq.~(\ref{v1}) is just the well-known Gromes relation, and Eq.~(\ref{v2})
gives a relation between the two extra spin-orbit coupling potentials
$\left( \dfr{{\rm\bf{S}_1 }}{m_1^2} -\dfr{{\rm\bf{S}_2 }}{m_2^2} \right)
{\bf \cdot L} \dfr{1}{r}\dfr{dV_5(r)}{dr}$ and
$\left( \dfr{{\rm\bf {S}_1 - {S}_2 }}{m_1m_2} \right)
{\bf \cdot L} \dfr{1}{r} \dfr{dV_6(r)}{dr} $, which agrees with the result in
Ref.\cite{4}. Our relations (\ref{g1})-(\ref{g4}) imply that the general
form of $V(r)$ given in Ref.\cite{4} is valid to all orders in perturbative
QCD in the second kind of approach, and even beyond perturbation.

In summary, we have derived, from reparameterization invariance, a set
of general relations between the SI and SD potentials of heavy quark and
antiquark interactions, namely eqs.~(\ref{i1})-(\ref{g4}). It includes the
Gromes relation (\ref{v1}) and a relation (\ref{v2}), which cover all the
results of the two kinds of approaches to the SD potentials\cite{1,20,2,3,4}.
Eqs.(\ref{i1}) and (\ref{i2}) are two new relations between various SI
potentials, which are useful for understanding the SI relativistic corrections
to the leading order static potential $V_0(r)$. Eq.(\ref{g4}) contains three
general restrictions to the form of the total potential showing that the three
terms containing $U_{11}(q)$, $U_{13}(q)$, and $U_{14}(q)$ in (\ref{mpot})
should actually vanish. Reparameterization invariance does not give any
constraints to the hyperfine and the tensor potentials.

We are grateful to S.-H. H. Tye for discussions.

\end{document}